# Interfacial activity of phosphonated-Polyethylene glycol functionalized cerium oxide nanoparticles

*L. Qi[a,c], J. Fresnais[b], P. Muller[a,d], O. Theodoly[a,e], J.-F. Berret[b] & J.-P. Chapel[a,f]\**

[a]*Complex Assemblies of Soft Matter Laboratory (COMPASS) - CNRS UMI3254 Rhodia Center for Research and Technology in Bristol - 350 Georges Patterson Blvd, Bristol, PA19007 - USA*
[b]*Matière et Systèmes Complexes (MSC), UMR 7057 CNRS/Université Denis Diderot Paris-VII Bâtiment Condorcet 10 rue Alice Domon et Léonie Duquet, 75205 Paris, France*
[c]*Lab of the Future (LOF), UMR 5258 Rhodia/CNRS/Université Bordeaux 1, 178 avenue du Docteur Schweitzer, F–33608 Pessac cedex, FRANCE*
[d]*Institut Charles Sadron, UPR 22 Université de Strasbourg, CNRS, 23 rue du Loess 67034 Strasbourg, France.*
[e]*Adhesion & Inflammation, INSERM U1067-CNRS UMR7333, and Université Aix-Marseille, Assistance Publique-Hôpitaux de Marseille, Case 937, 163 Avenue de Luminy, F-13009 Marseille, France.*
[f]*Univ. Bordeaux, Centre de Recherche Paul Pascal (CRPP), CNRS UPR 8641, F-33600 Pessac, France.*

chapel@crpp-bordeaux.cnrs.fr

**Abstract**

In a recent publication, we have highlighted the potential of phosphonic acid terminated PEG oligomers to functionalize strong UV absorption cerium oxide nanoparticles[1], which yield suspensions that are stable in aqueous or organic solvents and are redispersible in different solvent after freeze-drying. In the present work, we highlight the interfacial activity of the functional ceria nanoparticles and their potential to modify hydrophobic surfaces. We first investigated Phosphonated-PEG amphiphilic oligomers behavior as strong surface active species forming irreversibly adsorbed layers. We then show that the oligomers interfacial properties translate to the functional nanoparticles. In particular, the addition of a small fraction of phosphonated-PEG oligomers with an extra C16 aliphatic chain (*stickers*) into the formulation enabled the tuning of i) the nanoparticles adsorption at the air/water, polystyrene/water, oil/water interfaces and ii) the particle/particle interaction in aqueous solutions. We also found that dense and closely packed two dimensional monolayers of nanoceria can be formed by spontaneous adsorption or surface compression using a Langmuir trough. A hexagonal organization controlled by reversible and repulsive repulsion has been characterized by GISAXS. Mono- or multilayers can also be stably formed or transferred on solid surfaces. Our results are key features in the field of polymer surface modification, solid stabilized emulsions (Pickering) or supracolloidal assemblies.

***Keywords***: nanoparticle, cerium oxide, phosphonated-PEG, hydrophobic sticker, adsorption, interface, surface.

## Introduction

Engineered nanoparticles (NPs) made from noble metals, rare-earth oxides or semiconductors are certainly one of the key actors of the incoming nanotech developments and future applications. Interest stems from their size-dependent properties giving rise to unique physical features including high reactivity, electronic, magnetic or optical properties fueling innovations, and driving technological breakthroughs in many areas [2] from nanomedecine[3] to materials science[4]. Numerous synthesis routes of inorganic nanoparticles now exist in the literature and availability of small (<10 nm), fairly monodisperse and non-aggregated inorganic nanoparticles of various chemistries is no longer a limitation. Synthesized nanoparticles sols are however extremely sensitive to changes in their physico-chemical environment such as pH, ionic strength, temperature and concentration often leading to aggregation. This drawback has been addressed through the adsorption of an organic layer (corona, shell or adlayer) around the particle promoting an (electro)-steric stabilization offsetting the *van der Waals* interaction that drives the aggregation. Beyond the sole stabilization, the presence of specific groups at the surface can confer specific functionalities interesting for biomedical applications such as drug delivery, immunoassay or cell imaging, where the control of the interactions between nanoparticles and bio-macromolecules, and cells or living tissues drives toxicity [5-7].

In a recent publication[1] we have highlighted the importance and versatility of such organic shells made with phosphonic acid terminated PEG oligomers (PPEG) for the functionalization of rare earth cerium oxide nanoparticles (or nanoceria), which plays a growing role in science and technology ranging from material science (catalysis, polishing, optics…)[8-13] to biomedical[14-16] applications. The solvating brush-like phosphonated-PEG layer was sufficient to solubilize the particles and greatly expand the stability range up to pH 9. Furthermore, the tailored ceria sol not only maintained its original strong UV absorption capability but the presence of end-functional PPEG also created true redispersible nanopowders in aqueous solution and in certain organic solvents providing a framework for designing a truly versatile hybrid metal oxide sol.

A large body of work has been devoted during the last 30 years to the modification of liquid/liquid, liquid/air or liquid/solid interface using organic surfactants and macromolecules[17]. With the recent possibility to synthesize a large variety of inorganic nanoparticles with appealing properties, the interfacial behavior of such functional particles has attracted recently a lot of interest [4, 18, 19] as it enables the generation of smart interfaces combining the advantageous properties of both the organic world of *macromolecules* and the inorganic world of *nanoparticles* carrying their unique intrinsic properties. The presence of a specific ligand around the inorganic nanoparticles can enable furthermore the fine tuning of their interfacial self-assembly through adsorption driven by the minimization of the overall free energy of the system[19]. Such tailored *interfacial* activity can be advantageously used in the field of polymer surface modifications or functional coatings[4] via the formation of nanoparticles mono- or multilayers at the surface of a given substrate, supramolecular assembly[20] or formulation of solid-stabilized ''Pickering'' emulsions [13].

In this work we discuss the surface and interfacial activity of phosphonated-PEG functional ceria nanoparticles, which are a key player of the current nanotech development. In order to assess a good transfer of the interfacial properties of the oligomers to the functional ceria nanoparticles both adsorption behaviors at the (air/water), (solid/water) and (oil/water) interfaces were studied independently and compared using the rising bubble and optical reflectometry and cryo-TEM techniques. The formation of dense nanoparticle monolayers at the air/water interface of a Langmuir trough was studied via *insitu* GISAXs measurements. Their subsequent transfer onto a solid substrate as mono- or multilayer was also highlighted.

## Materials and Methods
### Nanoparticles and oligomers

**Nanoparticles:** The inorganic mineral oxide nanoparticle system investigated is a dispersion of cerium oxide nanocrystals, or nanoceria sol at pH 1.5. The synthetic procedure involves thermohydrolysis of an acidic solution of cerium-IV nitrate salt ($Ce^{4+}$ ($NO_3^-$)$_4$) at high temperature (70° C), that results in the homogeneous precipitation of a cerium oxide nanoparticle pulp ($CeO_2$ ($HNO_3$)$_{0.5}$ ($H_2O$)$_4$)[21]. The size of the particles was controlled by addition of hydroxide ions during the thermohydrolysis. High resolution transmission electron microscopy has shown that the nanoceria (bulk mass density ρ = 7.1 g/cm$^3$) consists of isotropic agglomerates of 2 to 5 crystallites with typical size of 2 nm and faceted morphologies. Wide-angle x-ray scattering has confirmed the crystalline fluorite structure of the nanocrystallites[22]. Image analysis performed on cryo-TEM images of single nanoparticles has shown a polydispersity index s = 0.15 ± 0.03 for the particles (s is defined as the ratio between the standard deviation and the average diameter)[23].

As synthesized, the cerium oxide nanosols are stabilized by combination of long range electrostatic forces and short range hydration interactions (including strongly bound or condensed nitrate ions). The hydrodynamic diameter is 8-10 nm, depending on the synthetic condition. At low pH the ionic strength arises from the residual nitrate counter-ions present in the solution and acidic protons. This ionic strength around 0.045 M gives a Debye screening length $K_D^{-1}$ ~1.5 nm. An increase of the pH (above 2) *or* of the ionic strength (>0.45 M) results in a reversible aggregation of the particles, and destabilization of the sols leading eventually to a macroscopic phase separation. For this system, the destabilization of the sols occurs well below the point of zero charge (pzc) of the ceria particles, pzc=7.9 [24]. The bare nanoceria particles at pH 1.5 have a zeta potential ζ = +30 mV and an estimated structural charge of $Q_{CeO_2}$ = +300 e. The charges are compensated by nitrate anions in the Stern and diffuse layers surrounding each particle.

**Oligomers:** The phosphonated-PEG oligomers are poly (ethoxyethylene) derivatives, differentiated by the number of EO units and the nature of the end group; they were produced by *Rhodia Inc.* The first has 10 EO units and is ended with an –H group; the second has 4 EO units and is ended with a –$CH_3$ group; the third one has 10 EO units and is ended with a -cetyl group -($CH_2$)$_{15}$-$CH_3$. They are named hereafter PPEG$_{10}$, PPEG$_4$ and hydrophobic sticker (or simply sticker) respectively. Their overall chemical structures are presented in Figure 1b. Titration curves with NaOH (1M) have shown the presence of two distinct pKa's for the phosphonated oligomers at

pKa$_1$ = 2.7 and pKa$_2$ = 7.8. A short poly (ethylene glycol) with 13 EO units and 2 –H end groups (M$_w$=586g/mol), named hereafter PEG, was used as a control in the evaluation of interfacial activity of oligomers. The PEG was purchased from Polysciences Inc. and used without further purification.

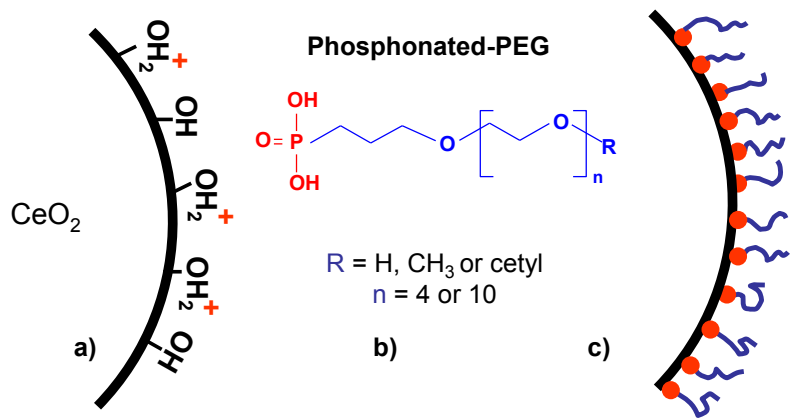

*Figure 1.* (a) Simplified sketch of the surface chemistry of cerium oxide nanoparticles: cationic protonated hydroxyl groups versus neutral hydroxyl groups. (b) Tailored phosphonated-PEG chemical architecture for the adsorption onto nanoparticles (c) Phosphonated-PEG corona around the particle: steric stabilization + PEG functionality.

The presence of a polar head and a relatively long aliphatic tail induces a self-assembly of stickers into micelles. The critical micellization concentration (CMC) was estimated at .07 mM by light scattering. This very low value hinders any efficient formulation at high concentrations with stickers still as unimers. We have thus artificially increased the CMC by adding ethanol in solution. CMC is known to vary as ~$e^{-\text{ f(geometry of amphiphilic structure)} \cdot \gamma}$[25], where $\gamma$ is the interfacial tension between the hydrophobic moieties and the solvent. In an ethanol/water (50/50 vol/vol) mixture, $\gamma$ dropped down to ~ 30mN/m, leading to a CMC around 7.3 mM (value obtained by conductivity measurements) that is 100 times higher than in pure water.

## Hybrid nanoparticles formulation

The reader is referred to our previous publication for more details [1]. The pH of the phosphonated-PEG (PPEG) solution is adjusted with reagent-grade nitric acid (HNO$_3$). Mixed solutions of CeO$_2$ nanoparticles and PPEG are prepared by simple mixing of dilute solutions prepared at the same weight concentration *c* (*c* = 0.1 - 1 wt. %) and the same pH. This ensures that no aggregation of nanoparticles occurs due to pH or ionic strength gap. At pH 1.4, 2.5 wt% of the phosphonate groups are ionized. The relative amount of each component is monitored by the volume ratio X between particles versus polymer solutions, yielding for the final concentrations:

$$c_{\text{CeO}_2} = \frac{cX}{1+X} \quad , \quad c_{\text{PPEG}} = \frac{c}{1+X} \tag{1}$$

In the case where hydrophobic stickers are used for the functionalization, the weight fraction of hydrophobic stickers out of all oligomers (PPEG and sticker) is fixed at either 1% or 5%, depending on the solvent nature, ensuring that the concentration of the stickers remains below their CMC. Ammonium hydroxide (NH$_4$OH) is then used to adjust the pH of CeO$_2$-PPEG dispersions between pH 1.5 and pH 10.

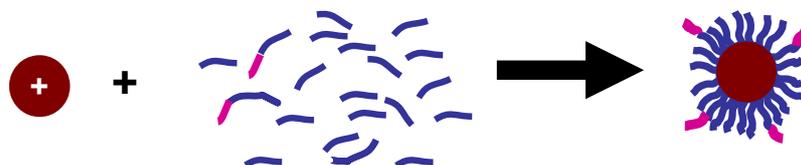

*Figure 2. Schematic representation of nanoceria functionalization with PPEG oligomers with a small fraction bearing a hydrophobic tail (stickers).*

Prior to adsorption experiments, the functional nanoparticles dispersions were dialyzed against pure water for 24 hours to remove any free oligomers still present in the bulk solution.

## Probing techniques

***Static and dynamic light scattering.*** Static and dynamic light scattering (SLS/DLS) measurements are performed on a BI-9000AT Brookhaven spectrometer (incident wavelength 488 nm). To accurately determine the size of the colloidal species, dynamic light scattering was performed with concentration ranging from c = 0.01 wt. % – 1 wt. %. In this range, the diffusion coefficient varies according to: $D(c) = D_0(1 + D_2c)$, where $D_0$ is the self-diffusion coefficient and $D_2$ is a virial coefficient of the series expansion. From the sign of the virial coefficient, the type of interactions between the aggregates, either repulsive or attractive can be deduced. A positive $D_2$ indicates a repulsive interparticle interaction. From the value of $D(c)$ extrapolated at c = 0 (noted $D_0$), the hydrodynamic radius of the colloids is calculated according to the Stokes-Einstein relation, $D_H = k_B T/3\pi\eta_S D_0$, where $k_B$ is the Boltzmann constant, T the temperature (T = 298 K) and $\eta_S$ ($\eta_S$ = 0.89×10$^{-3}$ Pa.s) the solvent viscosity. The autocorrelation functions of the scattered light are interpreted using the CONTIN fitting procedure.

***Optical reflectometry.*** The adsorption isotherms $\Gamma(c)$ of phosphonated-PEG oligomers onto a cerium oxide model surface is monitored using optical reflectometry [26, 27]. Fixed angle reflectometry measures the reflectance at the Brewster angle on the flat substrate. A linearly polarized light beam is reflected by the surface and subsequently split into a parallel and a perpendicular component using a polarizing beam splitter. As material adsorbs at the substrate-solution interface, the intensity ratio S between the parallel and perpendicular components of the reflected light varies. The change in S is related to the adsorbed amount through:

$$\Gamma(t) = \frac{1}{A_s} \frac{S(t) - S_0}{S_0} \quad (2)$$

where $S_0$ is the signal from the bare surface prior to adsorption. The sensitivity factor $A_S$, which is the relative change in the output signal S per unit surface, is found to be proportional to dn/dc. $\Gamma(c)$ is constructed by taking the plateau value $\Gamma_{Plateau}$ of a given $\Gamma(t)$ curve at different concentration c (c=0.001 to 0.1 wt. %).

*Adsorption*. Measurements were performed at the stagnation point of the flow; the adsorption kinetics is then not influenced by convection but only by diffusion and any existing energy barrier. The former is proportional to the bulk concentration, and the latter is related to the interaction between the upcoming molecule and the surface, as well as the interaction between the upcoming molecule and the already adsorbed molecules on the surface. At the beginning of the adsorption, the surface is unoccupied and the interaction between the upcoming molecule and the molecules on the surface can be neglected. Typically, we have (Fleer, Cohen Stuart et al. 1993)

$$\left(\frac{d\Gamma}{dt}\right)_{t=0} \propto c^{\beta} \qquad (3)$$

with $t$ the time, c the bulk concentration, and β is a parameter <1 that accounts for adsorption energy barrier effects. The case β = 1, encountered in case of high affinity and negligible barrier effects, corresponds to a diffusion controlled regime, for which the initial adsorption rate is proportional to the bulk concentration.

*$CeO_2$ model surfaces*. A thin layer of polystyrene (~ 100 nm) is deposited on top of an HMDS (hexamethyldisilizane) functionalized silicon wafer by spin-coating a toluene solution (25 g/l) at 5000 rpm. The surface is then dipped in a nanoceria solution (0.1 wt. %) containing 0.1 M $NaNO_3$ overnight. This results in the formation of a well packed nanoceria monolayer of thickness 8 nm on top of the PS surface. It should be noted that the receding water contact angle $\theta_r$ on such model nanoceria surface was below 15º in contrast with a $\theta_r$ =85 º for the original PS surface[28].

*Langmuir-Blodgett films*. A Teflon trough (Nima Technology Ltd., Coventry, U.K.) with a single barrier system was used. The surface pressure was measured by the Wilhelmy hanging-plate method. Chloroform solutions of both oligomers and functionalised nanoparticles (~ 50 µL, 1 mg/mL) were carefully spread drop by drop with the help of a microsyringe on the water surface of the trough. After the evaporation of the solvent, pressure–area isotherms (π-A) were recorded ($\Pi = \gamma_0 - \gamma$, where $\gamma_0$ is the surface tension in absence of a monolayer and $\gamma$ the surface tension with the monolayer and A the area occupied per molecule). The surface area occupied per molecule (A) can be calculated from the volume and concentration of deposited solution, trough geometry and molecular weight. In order to generate multilayer films, the monolayer was compressed to a given surface pressure and then transferred onto a solid substrate while keeping the pressure constant via a feed-back loop. In the case of a hydrophilic silicon wafers or glass slides the first layer is transferred by raising the substrate through the monolayer from the aqueous subphase. Multilayers can be achieved by successive rising/lowering of the solid substrate through the monolayer.

*Risingbubble*. The adsorption kinetics of oligomers or oligomer functionalized nanoparticles at the air/water interface was monitored by pendant drop measurements. When a hydrophobic zone exists in the structure of the solute, the molecule will tend to adsorb at the bubble/water interface. The subsequent increase of the surface concentration will lead to a decrease in the air/water surface tension. The experimental setup allows a real-time measurement of the surface pressure via the detection of the shape factors of the bubble. By measuring the surface pressure at saturation of solutions at different concentration, one obtains the surface-pressure isotherm, $\pi = f(c)$, where c is the concentration of molecules in solution[29]. The lag time $\tau_{lag}$ defined as the time interval between the generation of the bubble and the detection of a finite surface pressure was experimentally determined at the onset of the sharp increase of the surface pressure with time. The surface concentration at $\tau_{lag}$ can be evaluated from a Langmuir (trough) isotherm[30], and corresponds to the end of the transition zone between the gas and liquid state of the monolayer. In the case of the PPEG stickers at pH 6, $\Gamma_{lag}$ was found to be ~ 0.5 mg/m$^2$ for a $\pi_{lag}$ ~1mN/m.

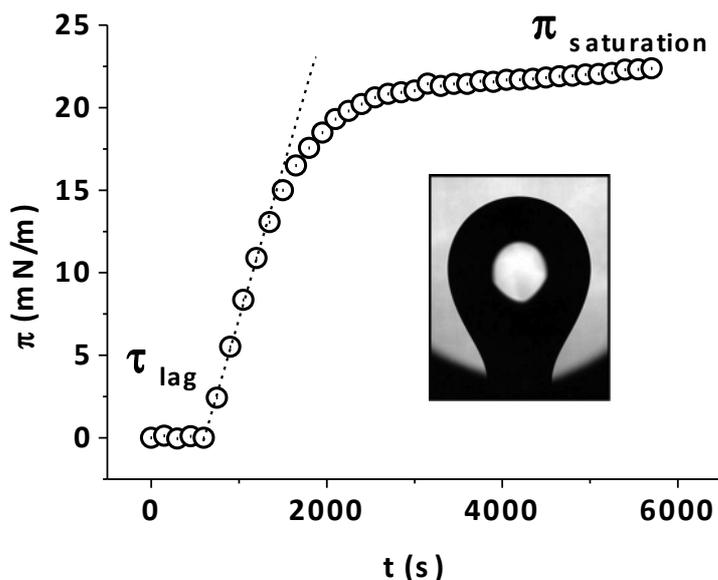

*Figure 3. Typical sticker adsorption curve at the air/water interface as monitored by pendant drop measurements. The lag time τ is found experimentally at the onset of the sharp increase of the surface pressure with time.*

***Grazing incident small angle X-ray scattering***. The lateral structure of a monolayer of PPEG functionalized nanoceria at the air/water interface was investigated by grazing incident small angle X-ray scattering (GISAXS). GISAXS was performed at the beamline X22B at the National Synchrotron Light Source, Brookhaven National Laboratory, using X rays with a wavelength of 1.567 Å. The exposure time was 10 s per frame and the incident angle was slightly smaller than the critical angle of the water subphase.

## Results and discussion

### Oligomers interfacial activity

We have in a first step investigated the adsorption behavior of the phosphonate-PEG oligomers alone at different interfaces in order to assess, in a second step, the efficiency of the interfacial *activity* ''transfer'' onto the inorganic ceria nanoparticles.

*Air/water interface*. Due to the amphiphilic nature of the oligomers, they are expected to adsorb at the air/water interface as regular surfactants do.

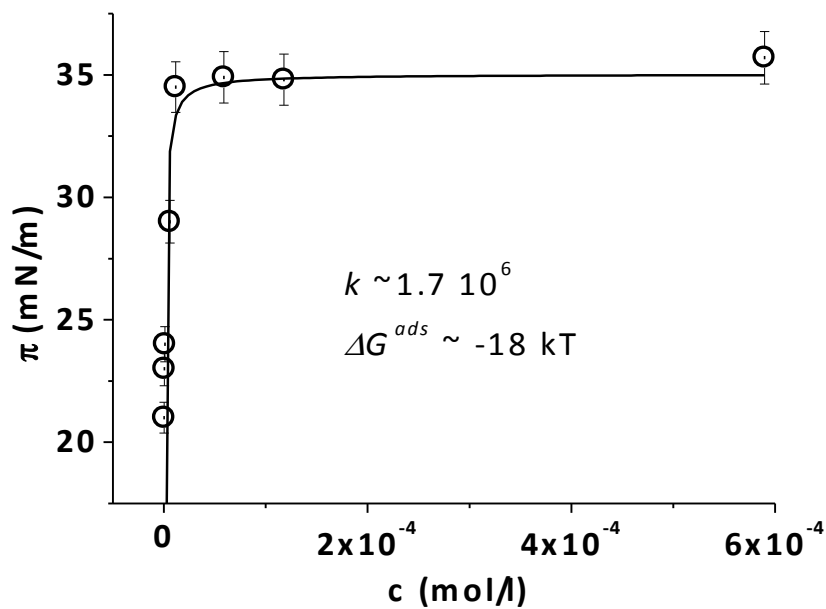

*Figure 4. Surface pressure Π as a function of sticker's concentration c in water. Solid line is a Langmuir fit (Eq. 4) to the data*

Figure 4 shows a typical oligomers isotherm with a pressure at saturation reached at very low concentration (c=$10^{-2}$ g/l) suggesting a strong affinity of the stickers with the air/water interface. The data were fitted using a Langmuir model where the sticker's interfacial pressure reads [1]

$$\pi(c) = \frac{k\pi_{sat}c}{1+kc} \tag{4}$$

where $c$ is the concentration of stickers in the aqueous solution, and $k$ is the adsorption constant ($k = k_{adsorption}/k_{desorption}$). The fit gives a pressure at saturation $\pi_{sat}$ = 35 mN/m and a large adsorption constant k~1.7x$10^6$ l/mol leading to small values for $k_{desorption}$, consistent with a high affinity type of isotherm. It should be noted that with the help of a Langmuir isotherm, $\pi_{sat}$ translates into an adsorbed amount of roughly 0.3 nm²/stickers or 4.5 mg/m². The free energy of adsorption $\Delta G^{ads}$ is estimated from the following expression

$$\Delta G^{ads} = -k_B T \ln(\frac{k}{V_{water}^m}) \tag{5}$$

where $k_B$ is the Boltzmann constant, $T$ is the temperature, and $V_{water}^m$ is the molar volume of water (0.018 L/mol). A $\Delta G^{ads}$ of -18 $k_B$T was deduced in line with what is found for non ionic ethoxylated surfactants at the air/water interface [31]. We have monitored via pendant drop measurements the stickers adsorption dynamics in order to sort diffusion-controlled and free convection adsorption regimes. In Figure 5, we can clearly observe that the lag time evolution exhibits a well-defined $c^{-2}$ behavior at high concentrations, then changes to a $c^{-1}$ behavior at low concentrations. The crossover between those two regimes occurs at $c^{crossover}$ ~ 4·$10^{-3}$% wt.

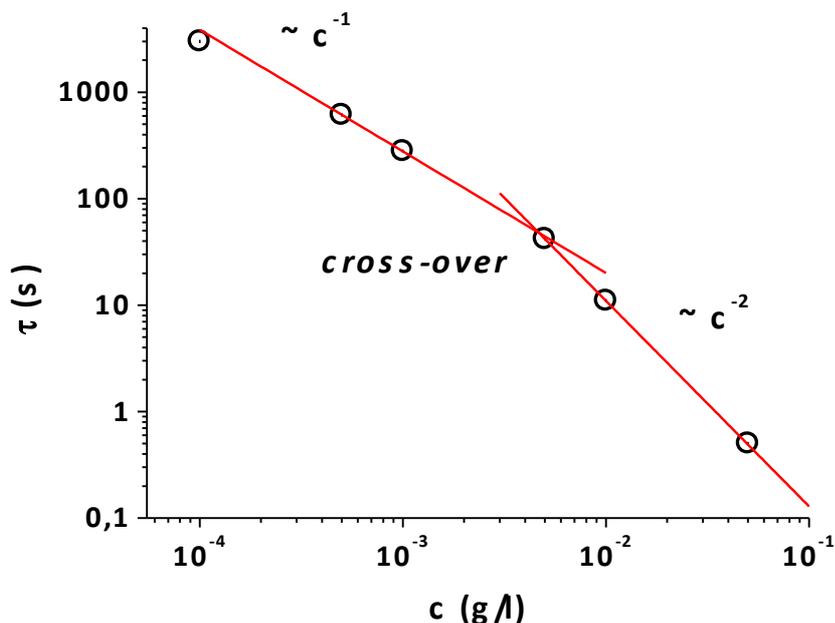

**Figure 5.** *Characteristic Lag time τ vs. concentration for stickers in water.*

Ybert *et al.* [29] have developed a model where *diffusion* and *natural convection* are both taken into account. In this model, the mass transport within a layer δ is governed by diffusion while free convection ensures that the concentration at the distance δ from the interface is always $c_0$, where $c_0$ is the bulk concentration. At small time scale a simple diffusion model in an infinite medium is recovered:

$$\tau_{lag} = \frac{\pi}{D}\left(\frac{\Gamma_{lag}}{2c}\right)^2 \propto c^{-2} \quad (6)$$

Whereas at large time scale the simple stationary layer model is found

$$\tau_{lag} = \frac{\Gamma_{lag} * \delta}{D * c} \propto c^{-1} \quad (7)$$

Within this model, the crossover between the two regimes occurs when molecules from distances greater than δ are required to reach $\Gamma_{lag}$; that is, when $D\tau \geq \delta^2$. With the value of $\Gamma_{lag}$ (~0.5 mg/m$^2$) measured from Langmuir trough isotherm (not shown here) and the crossover concentration $c^{crossover}$ (~ 0.005 g/l), we obtained a diffusion coefficient D ~1.25·10$^{-10}$ m$^2$/s, which is in excellent agreement with the value obtained from DLS measurements ( ~1.17·10$^{-10}$ m$^2$/s). With D and $c^{crossover}$, we can further deduce a $\tau^{crossover}$ of ~ 60 s and a diffusive layer width δ of ~ 83 μm.

We have then compared the adsorption of the three different oligomers at the air/water interface for a given bulk concentration (10$^{-3}$% wt). A clear difference in the rate of adsorption and the surface pressure at saturation was observed as seen in Figure 6. $\pi_{sat}$ increases with the increasing oligomer hydrophobicity, from PEGs, that are composed of hydrophilic EO units, to PPEGs that possess an hydrophobic propyl spacer between the phosphonate head and the EO units, and finally the stickers that have an extra hydrophobic aliphatic tail.

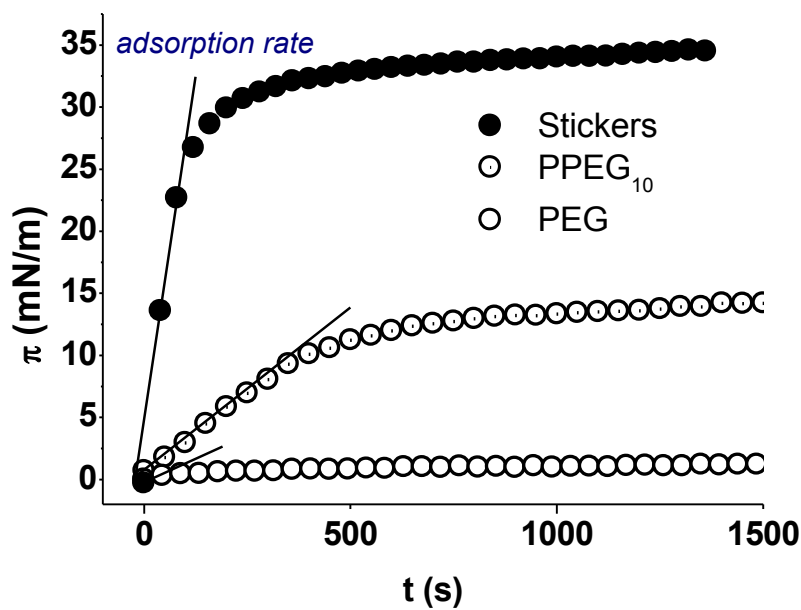

*Figure 6.* Adsorption behavior of PEG, PPEG$_{10}$ and stickers at the air/water interface. The curves show the evolution of surface pressure of a rising air bubble immersed into an oligomer solution (10$^{-2}$ g/l) as a function of time.

LB isotherms for PEG, PPEG and stickers are plotted in Figure 7. 20 µl of a chloroform solution was carefully spread drop by drop at the air/water interface. For the PEG oligomers, the surface pressure stays ~ 0mN/m up to an area per molecule A ~ 30 Å$^2$ where it starts to increase, indicating a gas-liquid transition in the monolayer. The surface pressure continues to increase until a plateau is finally reached at ~10 mN/m. At this stage a metastable equilibrium is reached and a further compression leads to the dissolution of some of the oligomers in the water subphase in agreement with the known poly(ethylene oxide) (PEO) air/water interfacial activity [32, 33]. For PPEG$_{10}$, the surface pressure increases more gradually due to the electrostatic repulsion between charged phosphonated headgroups (pH~5.6) before reaching a higher plateau at ~22 mN/m due to the presence of a hydrophobic propyl group in the backbone. A further compression leads here as well to the dissolution of the PPEG$_{10}$ in the subphase. In both cases, the molecules are partially soluble into the subphase hindering any accurate monitoring of the area per molecule.

In the case of the stickers, 10µL of solution is already enough to give rise to a surface pressure around 10mN/m due to the presence of hydrophobic cetyl tail. The surface pressure then increases regularly without the presence of a plateau until the collapse of the monolayer occurs around 40mN/m corresponding to an area per molecule of 25 Å$^2$ as in the case of insoluble fatty acids suggesting a strong anchoring of the stickers at the air/water interface.

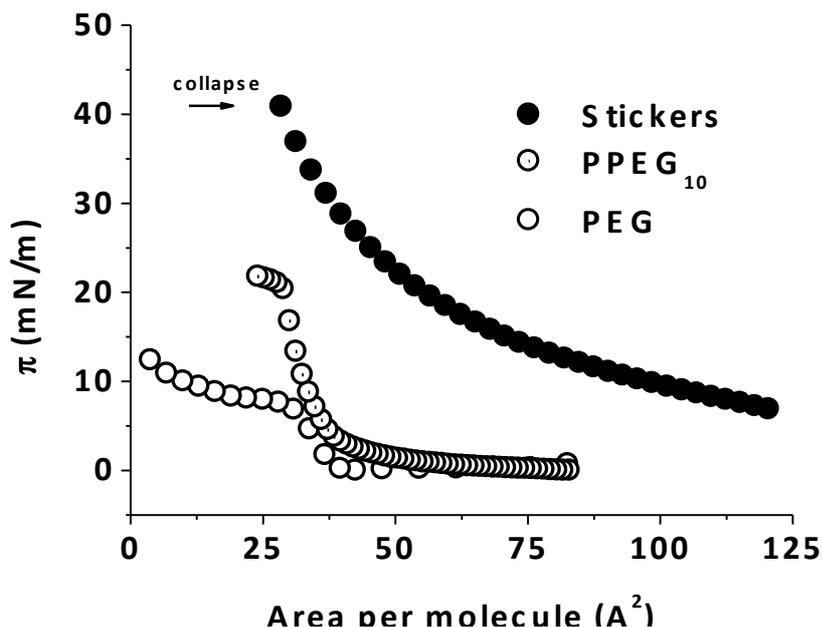

***Figure 7.*** *Langmuir-Blodgett isotherms for PEG, PPEG and stickers. The curves show the evolution of surface pressure Π with compression.*

***Hydrophobic substrate/water interface***. The affinity of both stickers and $PPEG_{10}$ oligomers towards hydrophobic surfaces was studied by monitoring their adsorption at 0.1 g/l onto a model polystyrene surface with an optical reflectometer. The saturated adsorbed amount was found 5 times lower for $PPEG_{10}$ (~0.28 mg/m$^2$) than for the stickers (~1.4 mg/m$^2$), which is consistent with a higher affinity of the hydrophobic cetyl tail for the hydrophobic PS surface. Furthermore, the adsorption isotherm built for the stickers. Figure 8 shows a sharp increase of the adsorption amount at low concentration typical of a high affinity isotherm. The data were fitted with a *Langmuir* model giving $\Gamma_{sat}$ = 1.38 mg/m$^2$, k = 4.2 10$^5$ l/mol, $\Delta G^{ads}$ = -17 $k_BT$. The strong affinity of stickers for the PS surface is of the same magnitude as for the air/water interface.

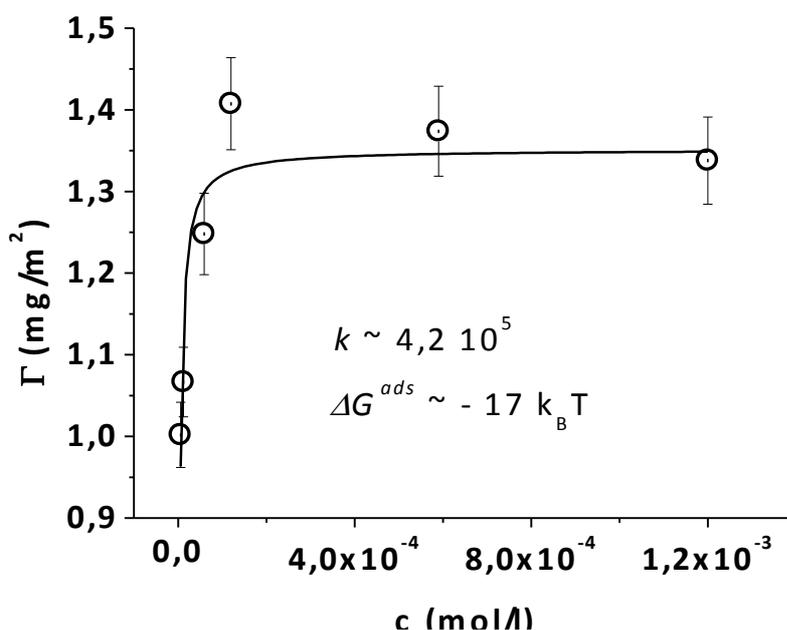

**Figure 8.** Adsorption isotherm of stickers onto a thin polystyrene layer spin-coated onto a silicon wafer.

Altogether, our results show that PPEG oligomers and stickers present a significant interfacial activity at the hydrophobic air/water interface with an enhanced affinity in case of the stickers as expected from their respective chemical structure. The next paragraph will investigate what will happen to the genuine oligomer interfacial activity when tethered around ceria nanoparticles.

**Interfacial properties of functional ceria nanoparticles**

*Air/water interface*. Rising bubble experiments performed on diluted (0.1% wt) aqueous solutions of bare and functionalized nanoparticles are presented on Figure 9 9. The bare charged $CeO_2$ nanoparticles do not adsorb to the air/water interface underlining the need of an organic functional corona. They are even slightly repelled from the interface as suggested by the small increase in the solution surface tension (visible on the red points in Figure 9 as a negative surface pressure), due to repulsive interactions between the charged bare nanoceria and their image charge (of same sign) at the air ($\varepsilon=1$)/water ($\varepsilon=80$) interface[34]. Remarkably, the presence of the PEG shell confers interfacial activity to the nanoparticles as can be seen in Figure 9. With oligomers, $\pi_{sat}$ increases with the hydrophobicity of the organic corona with the same $\pi_{sat}$=10 mN/m for free and tethered PPEGs. With just a small sticker fraction (1% wt) added during the formulation, $\pi_{sat}$ is further increased up to 22 mN/m, a value well between what was measured for both oligomers individually highlighting the possible fine tuning of the interfacial activity of the nanoparticles via the sticker content.

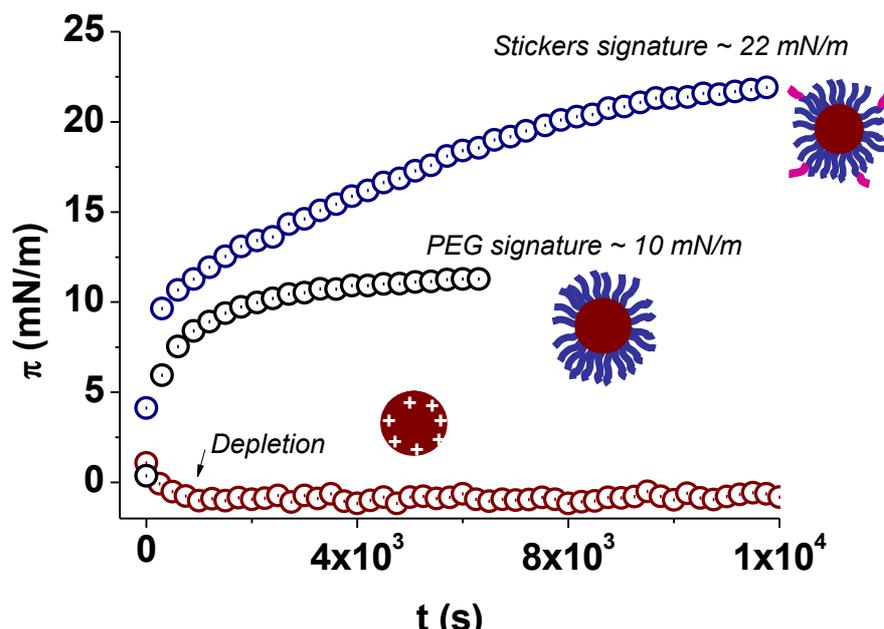

***Figure 9.*** *Evolution of the interfacial pressure Π of bare and functional $CeO_2$ nanoparticles with time monitored by rising drop measurements.*

Langmuir compression isotherms of $CeO_2$-$PPEG_{10}$/stickers (99/1=wt/wt), $CeO_2$-$PPEG_{10}$ and $CeO_2$-$PPEG_4$ monolayers, measured using a Langmuir trough, are plotted in Figure 10. The $CeO_2$-$PPEG_4$ shows a steep slope whereas the $CeO_2$-$PPEG_{10}$ is characterized by an early repulsive regime and a smaller surface pressure at low surface area possibly due to the presence of residual charges on the oxide core (as suggested by zeta potential measurements not shown here). In the case of the $CeO_2$-PPEG/stickers, the monolayer undergoes a plateau around 10mN/m indicative of the presence of ethoxylated blocks in the organic functional layer, then rises up steeply due to the close-packing of the nanoparticles bearing some cetyl tails in the corona. The position of the limiting areas where the pressure increases steeply ranges between 5000 Å$^2$ and 7500 Å$^2$ giving equivalent sphere radii between 4 and 5 nm in good agreement with dynamic light scattering measurements[Qi, Sehgal et al. 2008]. Hence, our results demonstrate the possibility to assemble nanoceria into two-dimensional monolayers of densely packed particles in repulsive and reversible interactions. It is interesting to question the internal structure of these assemblies.

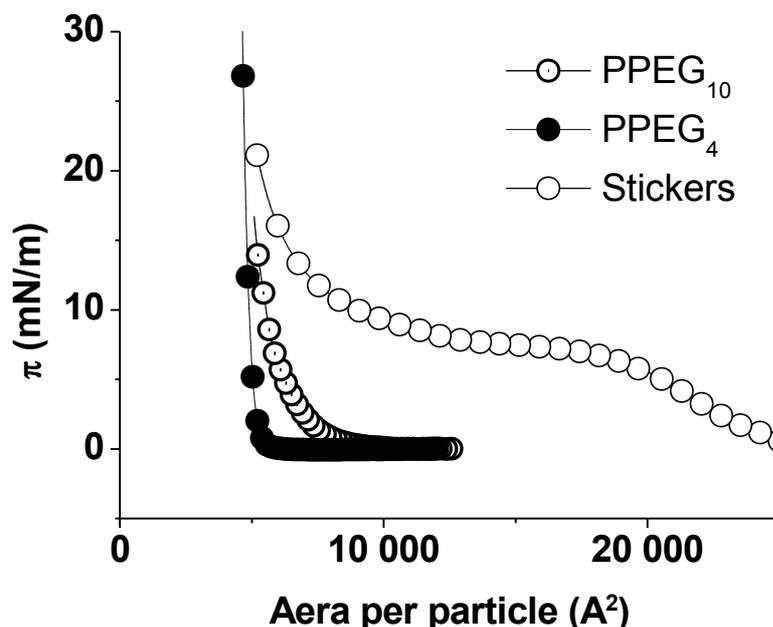

**Figure 10.** *LB compression isotherms for $CeO_2$-$PPEG_{10}$/stickers (99/1=wt/wt), $CeO_2$-$PPEG_{10}$ and $CeO_2$-$PPEG_4$. A collapse of the monolayer is observed above 40 mN/m in the case of the $CeO_2$-$PPEG_{10}$/stickers as in the case of the stickers alone (Figure 7).*

In order to describe the particle arrangement at the air/water interface GISAXS measurements have been performed *in-situ* on a monolayer of $CeO_2$-$PPEG_4$ with the help of a LB trough (**Error! Reference source not found.**)[35]. At any area for which the surface pressure is higher than 0.5 mN/m, the GISAXS data exhibits the characteristic pattern of a 2D hexagonal symmetry (as shown by the presence of a second order peak at √3 times the first order peak position). From the position of the first peak $Q_{//}$, assuming a hexagonal packing, the distance d* between nanoparticles can be calculated $\frac{4\pi}{\sqrt{3}}\frac{1}{Q_{//}}$. d* follows a density rule d* ~ √A, where A is the surface area.
This result is consistent with a 2D dilution law of a monolayer of individual particles with repulsive interactions. This rule is broken at close contact (8 nm), corresponding to the nanoparticles diameter in a hexagonal close-packed structure in very good agreement with light scattering data. Altogether, we have shown that functionalized nanoceria anchor strongly at a hydrophobic air/water interface and form two-dimensional layers with tunable packing. A particularly interesting application would consist in transferring such layers onto solid surface to allow production of new and tunable functional coatings.

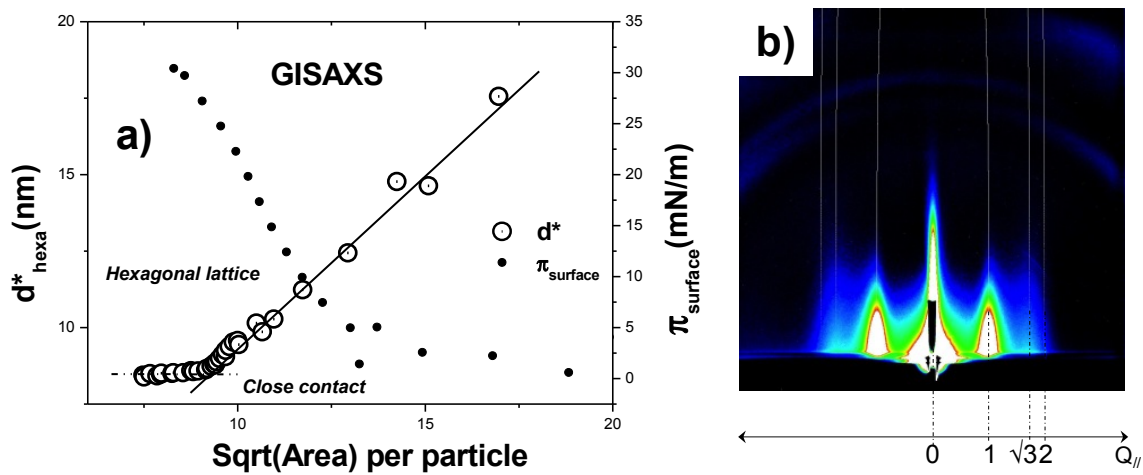

***Figure 11.*** *a) Evolution of the nanoparticle spacing d\* (left axis) as a function of the square root of surface area per particle of a monolayer of $CeO_2$-$PPEG_4$ monitored by GISAXS with corresponding surface pressure (right axis). b) 2D scattering patterns demonstrating the hexagonal symmetry.*

*Multilayer formation.* Due to the good *anchoring* of the functional ceria nanoparticles at the air/water interface, dense monolayers are formed at the water surface of a Langmuir trough, prior to the transfer onto solid substrates. On a *hydrophilic* silicon wafer surface, the different layers are deposited during the raising cycle only in order to minimize the interfacial energy as displayed on Figure 12. On hydrophobic substrates (silanised silicon wafer or PS thin film), the first monolayer is deposited when the sample is lowered into the water subphase; the followings during the raising step. In the case of $CeO_2$-$PPEG_4$ nanoparticles, the multilayer was homogeneous as shown by AFM measurements at least up to 6 sequential depositions. This approach enables the growth of rather thick (~50-100 nm) *all-nanoparticles* layer made without any intermediates as encountered in the well known layer *by* layer approach[2] where a given build-blocks (nanoparticles, polyelectrolytes etc..) is grown using an oppositely charged macromolecular binders.

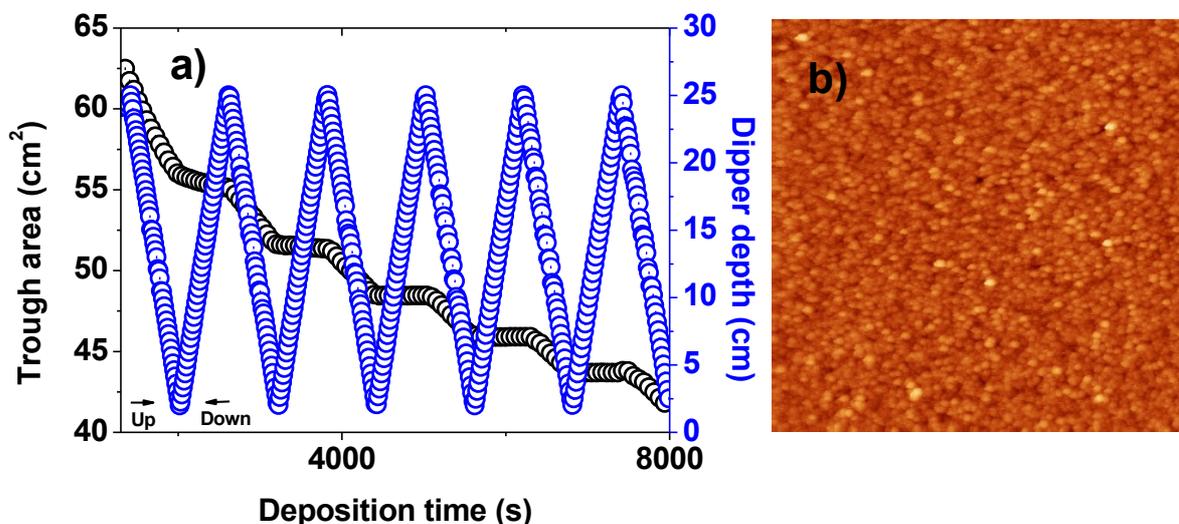

*Figure 12.* a) Multilayer made out of PPEG$_4$-functionalized nanoceria via the Langmuir-Blodgett deposition technique at a surface pressure of 20 mN/m. The integrity of the monolayer transfer was ascertained by the variation of the trough area during deposition (blue circles: dipper height, black circles: trough area). b) AFM image of silicon wafer surface covered with 6 layers of functional nanoceria particles (image size 1x1 µm) with an average monolayer thickness of 6 nm (from ellipsometric measurements.)

***Oil/water interface.*** Since their discovery at the beginning of the 20$^{th}$ century by Pickering the solid stabilized emulsions have attracted much attention. These so called *Pickering emulsions* are formed by the self-assembly of colloidal particles, either nano- or micro-particles, at fluid–fluid interfaces in two-phase liquid systems[19]; the wettability of the colloids at the oil−water interface determines the direct (oil-in-water) or inverse (water-in-oil) nature of the emulsion. We used the less surface active PPEG$_{10}$ functional ceria nanoparticles to highlight their affinity with oil/water interface. As shown in Figure 13, these nanoparticles are good Pickering emulsion agents inhibiting the coalescence of water/hexadecane emulsion droplets for at least 3 months.

In the case of Pickering emulsion stabilized by microscopic particles, the decrease of the overall free energy upon adsorption of a particle at the oil/water interface is much larger than thermal energy (a few k$_B$T) of the particle, leading to an effective confinement of large colloids to the interface. Conversely, the confinement of nanoparticles to the oil/water interface due to surface energy reduction is comparable to thermal energy[13]. Hence, nanoparticles display a dynamic exchange between the interface and the bulk, enabling nanoparticle assembly at the interface to attain its equilibrium[19] as for simple surfactants. The PPEG$_{10}$-CeO$_2$ nanoparticles do not exhibit an high affinity with the air/water (or the oil/water) interface enabling then a dynamic exchange with the bulk solution.  Cryo-TEM images were taken 3 months after the emulsion preparation as seen in the right bottom of Figure 13 where one can see an oil droplet surrounded by individual ceria nanoparticles (small dark spots of ~8nm).

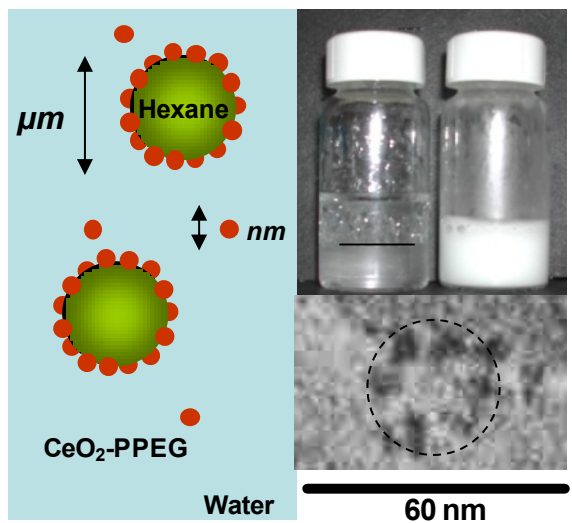

**Figure 13.** Pickering emulsion of water/hexadecane obtained by adding PPEG$_{10}$ functionalized nanoceria. Top right: non stabilized (phase separation) and stabilized water/hexane mixture. Bottom right: Cryo-TEM image of a hexane droplet from a 3-month-old-prepared emulsion stabilized by functional nanoceria (small dark spots of ~8nm along the dotted circular line).

***Hydrophobic substrate/water interface***. As in the case of individual oligomers, the affinity of phosphonated-PEG functionalized nanoceria for hydrophobic polymer surfaces was further evaluated by optical reflectometry. The adsorption rate at t=0 for both CeO$_2$-PPEG and CeO$_2$-PPEG/stickers (at 99/1 weight ratio) nanoparticles were plotted as a function of concentration (Figure 14). Linear scaling was found for both functional nanoparticles, indicating a diffusion controlled kinetics ($\beta$=1) suggesting a strong affinity with the surface, which is similar to the behavior at the air/water interface. The slight difference observed between the 2 slopes might be attributed to a different size and consequently to a different diffusion coefficient between the 2 functional nanoparticles.

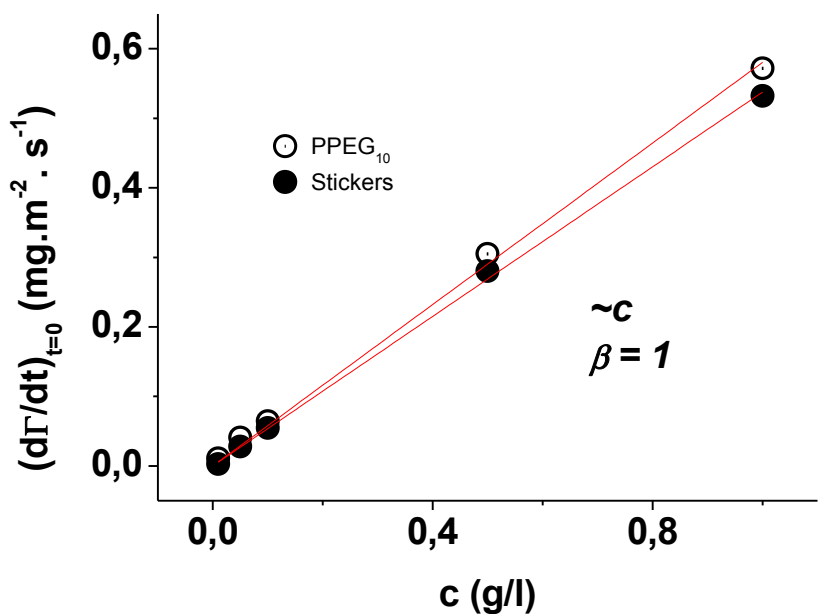

**Figure 14.** *Variation of the initial adsorption rate of $CeO_2$-$PPEG_{10}$ and $CeO_2$-$PPEG_{10}$/stickers (99/1=wt/wt) onto PS surface as a function of the concentration monitored by optical reflectometry.*

Adsorption isotherms were constructed (Figure 15) to characterize further the affinity of functionalized nanoceria for hydrophobic solid surfaces. A fit using a Langmuir model yields $\Gamma_{sat}$ = 7.4 mg/m², k = 5.5·10⁷ l/mol, and $\Gamma_{sat}$ =11 mg/m², k = 2·10⁷ l/mol for $CeO_2$-$PPEG_{10}$ and $CeO_2$-$PPEG_{10}$/stickers (99/1=wt/wt) respectively. The high values found for k confirm a strong affinity for the surface. It should be mentioned that an estimation of the adsorbed amount of a dense monolayer of functional nanoparticles (2D close-compact hexagonal arrangement) gives between 7-10 mg/m² in very good agreement with the experimental result (the molecular weight of the functional particles $M_w^{FNP}$ = 439 000 g/mol was taken from our previous work [1]). The free adsorption energy values were estimated to $\Delta G^{ads}$ = -20 $k_BT$ and $\Delta G^{ads}$ = -22 $k_BT$ respectively. It should be underlined that the presence of such a small amount of stickers (~1%) in the shell translates in a noticeably higher adsorbed amount (1.5 times higher) and higher adsorption energy (~10%), which proves that stickers concentration can be used to efficiently tune the final adsorption amount on different low energy surfaces via the sticker's content. After adsorption of the functional particles, the hydrophobic PS surface (contact angle with water θ~ 90°) turned hydrophilic (water sheeting effect with θ < 30°) and less sensitive to UV light due the known UV adsorption capability of the ceria nanoparticles[1]; a very interesting feature in the field of polymer surface modification and ageing.

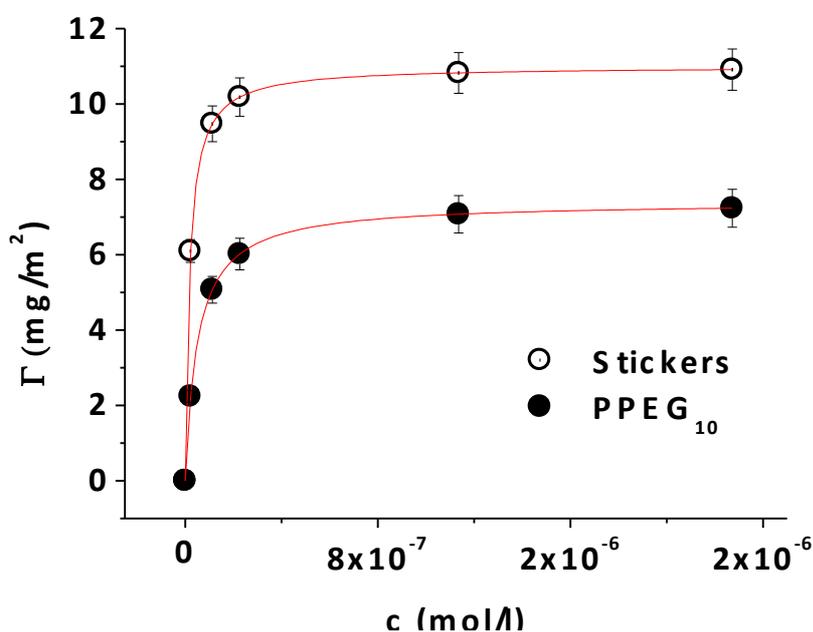

**Figure 15.** *Adsorption isotherms of $CeO_2$-$PPEG_{10}$ and $CeO_2$-$PPEG_{10}$/stickers (99/1=wt/wt) onto PS surface monitored by optical reflectometry. Red lines are Langmuir fit to the data.*

***Mixed solvent***. After assessing the affinity of the functional nanoparticles with an external interface, we have investigated the inter-particle interaction of sticker's functionalized nanoceria with the solution surface tension. The influence of the $PPEG_{10}$/stickers weight ratio on the interaction between functional particles was first investigated. Two batches of hybrid nanoparticles were formulated at X=0.5 and 1% wt. in an ethanol/water (E/W) vol/vol=50/50 mixture. The mixture composition was chosen to vary the sticker fraction on a large scale without forming any micelle. Weight ratios of 99/1 and 95/5 were investigated. In both cases, the particle size distribution was narrow with distinct $R_h$ values (6.0± 3.0 nm and 11.8± 6.5 respectively). At higher sticker content, due to a stronger hydrophobic interaction between the nanoparticles, doublet and triplet are formed. The influence of the ethanol/water volume ratio on the interaction between functionalized particles was then studied for the 95/5 ratio. E/W ratios of 50/50 vol/vol and 40/60 vol/vol were used. Both solutions were then diluted with the corresponding mixed solvent to different nanoparticles concentrations (0.3%, 0.5% and 0.75% wt.). DLS measurements were then performed and the mutual diffusion coefficient D was calculated. For both series, D varies linearly with the concentration c in agreement with $D(c) = D_0 (1 + D_2 c)$, as shown in Figure 16. At higher ethanol content (E/W=50/50), the virial coefficient $D_2$ (~ slope of D vs. c) is positive ($1.09 \cdot 10^{-9}$ cm$^2$.s$^{-1}$.L.g$^{-1}$) indicating a slight repulsion between the particles; at lower ethanol content (E/W=40/60), $D_2$ becomes negative ($-4.9 \cdot 10^{-9}$ cm$^2$.s$^{-1}$.L.g$^{-1}$) indicating a slight attraction. In the latter case the steric repulsion provided by the PEG chains is balanced by a larger hydrophobic (attractive) interaction provided by the stickers. From the value of D(c) extrapolated to c = 0, we can deduce the self-diffusion coefficients $D_{01}=1.08 \cdot 10^{-7}$ cm$^2$/s and $D_{02}=6.92 \cdot 10^{-8}$ cm$^2$/s. The hydrodynamic radii of the colloids were then found to be $R_{H01}$=8.6nm and $R_{H02}$=14.7nm in an E/W=40/60 and E/W=50/50 solution respectively. The difference in $R_H$ is due to the difference in the solvent quality. In a relatively bad solvent (E/W=40/60), the organic adlayer of the hybrid particles tends to collapse and give an apparently smaller hydrodynamic radius (the viscosity of the mixture has been taken into account). The presence of hydrophobic stickers around the ceria nanoparticles further enabled the fine tuning of the inter-particle interaction through the sticker fraction or the solvent surface tension offering a simple method to create hybrid supracolloidals assemblies of inorganic nanoparticles of different nature.

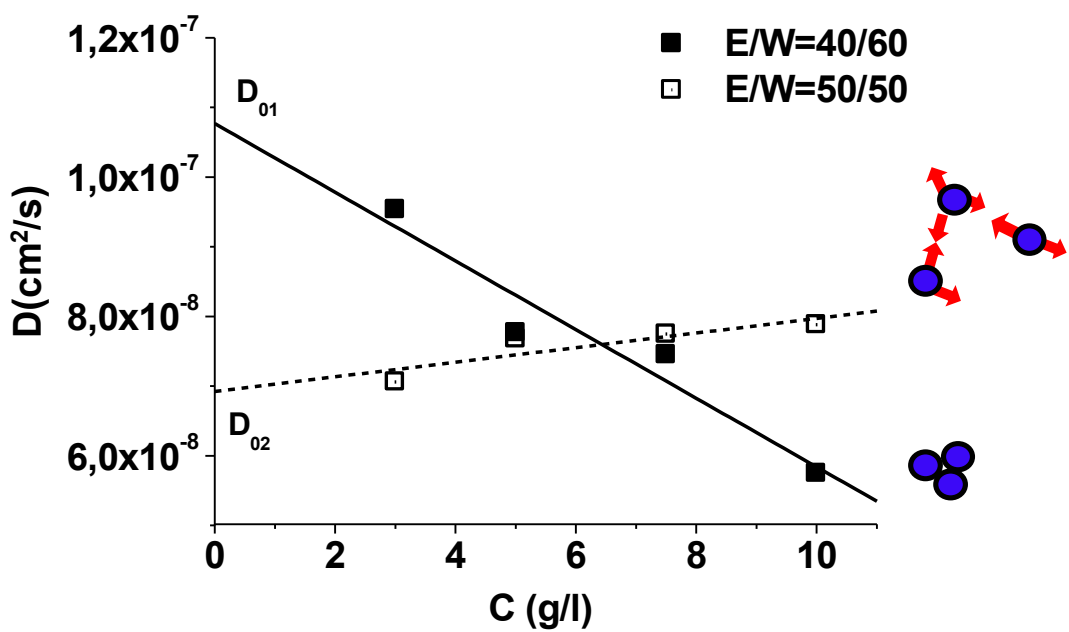

*Figure 16. Mutual diffusion coefficient D vs. nanoparticle concentration c for nanoceria functionalized with PPEG/ stickers (95/5=wt/wt) in ethanol/water mixtures.*

## Concluding Remarks

The interfacial properties of phosphonated-PEG cerium oxide nanoparticles were highlighted through a detailed series of complementary experiments. Both PPEG oligomers and stickers showed an increasing surface activity as shown by their controlled adsorption onto air/water, polystyrene/water and oil/water interfaces. This interfacial affinity was transferred to ceria nanoparticles functionalized by these molecules. After functionalization, the particles were indeed able to form dense monolayers at the air/water interface triggering the growth of thick hybrid layers via the sequential transfer onto hydrophilic or hydrophobic solid substrates. They were shown to be good Pickering emulsion agents able to stabilize oil droplets in water. Their high affinity towards polymer substrates turned the initially hydrophobic and UV sensitive surface to a hydrophilic and UV resistant surface. Furthermore, by tuning both the stickers' fraction and the solvent surface tension during the formulation, it was possible to change the inter-nanoparticles interaction, opening pathways for controlled supercolloidal assemblies. Finally, the effective interfacial properties of the phosphonated-PEG (+ stickers) adlayer, together with the intrinsic properties of the mineral oxide core (anti-UV in the case of $CeO_2$), will make these hybrid functional rare-earth nanoparticles excellent coatings, wetting or adhesion agents. Beyond, this *robust functionalization* platform extendable to other types of nanoparticles enables the translation of intrinsic properties of mineral oxide nanoparticles to various critical *bulk* and *surface* end uses.

## Acknowledgements.

The authors wish to thank the *Rhodia chemical company* for its financial support and Annie Vacher (CRTA, Aubervilliers, France) for performing cryo-TEM experiments. A special thanks to Jean-Christophe Castaing for fruitful discussions and encouragements over the years.

## References

1. Qi, L., et al., *Redispersible Hybrid Nanopowders: Cerium Oxide Nanoparticle Complexes with Phosphonated-PEG Oligomers.* ACS Nano, 2008. **2**(5): p. 879-888.
2. Chapel, J.P. and J.F. Berret, *Versatile electrostatic assembly of nanoparticles and polyelectrolytes: Coating, clustering and layer-by-layer processes.* Current Opinion in Colloid & Interface Science, 2012. **17**(2): p. 97-105.
3. Kim, J., Y. Piao, and T. Hyeon, *Multifunctional nanostructured materials for multimodal imaging, and simultaneous imaging and therapy.* Chemical Society Reviews, 2009. **38**(2): p. 372-390.
4. Nie, Z., A. Petukhova, and E. Kumacheva, *Properties and emerging applications of self-assembled structures made from inorganic nanoparticles.* Nat Nano, 2010. **5**(1): p. 15-25.

5. Rabin, O., et al., *An X-ray computed tomography imaging agent based on long-circulating bismuth sulphide nanoparticles.* Nat. Mater., 2006. **5**(2): p. 118-122.
6. Liu, G.L., et al., *A nanoplasmonic molecular ruler for measuring nuclease activity and DNA footprinting.* Nat. Nanotechnol., 2006. **1**(1): p. 47-52.
7. Bertorelle, F., et al., *Fluorescence-modified superparamagnetic nanoparticles: Intracellular uptake and use in cellular imaging.* Langmuir, 2006. **22**(12): p. 5385-5391.
8. Yu, T.Z., et al., *Ultraviolet electroluminescence from organic light-emitting diode with cerium(III)-crown ether complex.* Solid-State Electron., 2007. **51**(6): p. 894-899.
9. Hung, I.M., et al., *Preparation of mesoporous cerium oxide templated by tri-block copolymer for solid oxide fuel cell.* Electrochim. Acta, 2004. **50**(2-3): p. 745-748.
10. Patsalas, P., S. Logothetidis, and C. Metaxa, *Optical performance of nanocrystalline transparent Ceria films.* Appl. Phys. Lett., 2002. **81**(3): p. 466-468.
11. Feng, X.D., et al., *Converting ceria polyhedral nanoparticles into single-crystal nanospheres.* Science, 2006. **312**(5779): p. 1504-1508.
12. Suphantharida, P. and K. Osseo-Asare, *Cerium oxide slurries in CMP. Electrophoretic mobility and adsorption investigations of ceria/sificate interaction.* J. Electrochem. Soc., 2004. **151**(10): p. G658-G662.
13. Zgheib, N., et al., *Stabilization of Miniemulsion Droplets by Cerium Oxide Nanoparticles: A Step toward the Elaboration of Armored Composite Latexes.* Langmuir, 2012. **28**(14): p. 6163-6174.
14. Tarnuzzer, R.W., et al., *Vacancy engineered ceria nanostructures for protection from radiation-induced cellular damage.* Nano Lett., 2005. **5**(12): p. 2573-2577.
15. Chen, J.P., et al., *Rare earth nanoparticles prevent retinal degeneration induced by intracellular peroxides.* Nat. Nanotechnol., 2006. **1**(2): p. 142-150.
16. Schubert, D., et al., *Cerium and yttrium oxide nanoparticles are neuroprotective.* Biochem. Biophys. Res. Commun., 2006. **342**(1): p. 86-91.
17. Lindman, B., ed. *Surfactants and Macromolecules : Self-Assembly at Interfaces and in Bulk.* 1991, Springer Verlag.
18. Grzelczak, M., et al., *Directed Self-Assembly of Nanoparticles.* ACS Nano, 2010. **4**(7): p. 3591-3605.
19. Boker, A., et al., *Self-assembly of nanoparticles at interfaces.* Soft Matter, 2007. **3**: p. 1231-1248.
20. Boal, A.K., et al., *Self-assembly of nanoparticles into structured spherical and network aggregates.* Nature, 2000. **404**(6779): p. 746-748.
21. Chane-Ching, J.-Y., *Preparing a dispersible, sol-forming cerium (IV) composition.* U.S. Patent 5308548 A, 1994.
22. Chen, H.I. and H.Y. Chang, *Synthesis of nanocrystalline cerium oxide particles by the precipitation method.* Ceram. Int., 2005. **31**(6): p. 795-802.
23. Berret, J.F., et al., *Stable oxide nanoparticle clusters obtained by complexation.* J. Colloid Interface Sci., 2006. **303**(1): p. 315-318.
24. M. Nabavi, O.S.a.B.C., *Surface Chemistry of Nanometric Ceria Particles in Aqueous Dispersions.* J. Colloid Interface Sci., 1993. **160**: p. 459 - 471.

25. Israelachvili, J.N., *Intermolecular and Surface Forces* London: Academic Press, First ed. (1985), Second ed. (1991)

26. Fleer, G.J., et al., *Polymers at Interfaces*. 1993, London: Chapman & Hall.
27. Dijt, J.C., M.A.C. Stuart, and G.J. Fleer, *Reflectometry as a tool for adsorption studies.* Adv. Colloid Interface Sci., 1994. **50**: p. 79 – 101.
28. Chapel, J.-P., A. Rao, et al., *Easy route to surface functionalization using charged inorganic nanoparticles.* In preparation, 2009.
29. Christophe Ybert and J.-M.d. Meglio, *Study of Protein Adsorption by Dynamic Surface Tension Measurements: Diffusive Regime.* Langmuir, 1998. **14**: p. 471-475.
30. Théodoly, O., et al., *Adsorption Kinetics of Amphiphilic Diblock Copolymers: From Kinetically Frozen Colloids to Macrosurfactants.* Langmuir, 2009. **25**(2): p. 781-793.
31. Mishra, S.K. and D. Panda, *Studies on the adsorption of Brij-35 and CTAB at the coal–water interface.* J. Colloid Interface Sci., 2005. **283**(2): p. 294-299.
32. Walker, Y.C.a.L.M., *Surface tension driven jet break up of strain-hardening polymer solutions* Journal of Non-Newtonian Fluid Mechanics, 2001. **100**(1-3): p. 9-26.
33. Barentin, C., P. Muller, and J.F. Joanny, *Polymer Brushes Formed by End-Capped Poly(ethylene oxide) (PEO) at the Air−Water Interface.* Macromolecules, 1998. **31**(7): p. 2198-2211.
34. Netz, R.R., *Charge regulation of weak polyelectrolytes at low- and high-dielectric-constant substrates.* J. Phys.: Condens. Matter, 2003. **15** p. 239-244.
35. Theodoly, O., A. Checco, and P. Muller, *Charged diblock copolymers at interfaces: Micelle dissociation upon compression.* EPL (Europhysics Letters), 2010. **90**(2): p. 28004.